\documentclass[conference,a4paper,final,twocolumn,10pt,twoside]{IEEEtran}
\IEEEoverridecommandlockouts 

\ifCLASSINFOpdf

\else
\fi

\usepackage{mathtools}
\usepackage{amsmath}
\usepackage{amssymb}
\usepackage{cite}
\usepackage{graphicx}
\usepackage{epstopdf}
\usepackage{siunitx}
\DeclareMathOperator{\E}{\mathbb{E}}
\begin{document}
%
\title{On the BER of Multiple-Input Multiple-Output Underwater Wireless Optical Communication Systems}
\author{\IEEEauthorblockN{Mohammad Vahid Jamali, and Jawad A. Salehi}
\IEEEauthorblockA{Electrical Engineering Department, Sharif University of Technology, Tehran, Iran\\Email: m\_v\_jamali@ee.sharif.edu, jasalehi@sharif.edu
}}



\maketitle
\begin{abstract}
In this paper we analyze and investigate the bit error rate (BER) performance of multiple-input multiple-output underwater wireless optical communication (MIMO-UWOC) systems. In addition to exact BER expressions, we also obtain an upper bound on the system BER. To effectively estimate the BER expressions, we use Gauss-Hermite quadrature formula as well as approximation to the sum of log-normal random variables.
 We confirm the accuracy of our analytical expressions by evaluating the BER through photon-counting approach.
  Our simulation results show that MIMO technique can mitigate the channel turbulence-induced fading and consequently, can partially extend the viable communication range, especially for channels with stronger turbulence.
\end{abstract}
\IEEEpeerreviewmaketitle

\section{Introduction}
Underwater wireless optical communication (UWOC) has been recently introduced to meet requirements of high throughput and large data underwater communications. Acoustic communication systems, which have been investigated and implemented in the past decades have some impediments which hamper on their widespread usage for today's underwater communications. In other words, UWOC systems have larger bandwidth, lower latency and higher security than acoustic communication systems \cite{tang2014impulse}. These unique features suggest UWOC system as a desirable alternative to acoustic communication systems.

Despite of all the interesting specifications of UWOC systems, they are only suitable for low range underwater communications, i.e., typically less than $100$ m. This is mainly due to the severe absorption, scattering and turbulence effects of underwater optical channels. Absorption and scattering of photons through propagation under water cause attenuation and time spreading of the received optical signals \cite{tang2014impulse,gabriel2013monte,cox2012simulation,mobley1994light}. On the other hand, underwater optical turbulence which is mainly due to the random variations of refractive index (because of salinity and temperature fluctuations) results in fading of the propagating optical signal \cite{tang2013temporal,korotkova2012light}.

Prior works mainly focused on the study of absorption and scattering effects of UWOC channels. In \cite{tang2014impulse,gabriel2013monte,cox2012simulation} the channel turbulence-free impulse response has been simulated and modeled using Monte Carlo simulation method. In \cite{akhoundi2015cellular}, a cellular topology for UWOC network has been proposed and also the uplink and downlink BERs of such a network with optical code division multiple access (OCDMA) technique have been investigated. On the other hand, some useful studies have been accomplished to characterize and investigate turbulence effects of UWOC channels. For examples, in \cite{korotkova2012light} the scintillation index of optical plane and spherical waves propagating in weak oceanic turbulence channel, has been evaluated using Rytov method. Also the average BER of an UWOC system with log-normal fading channel has been investigated in \cite{yi2015underwater,gerccekciouglu2014bit}.
  Moreover, beneficial application of multi-hop transmission on the performance of underwater wireless OCDMA networks has been investigated in \cite{jamali2015performance-relay}.

In this paper we analytically study the BER performance of an UWOC system, with respect to the all impairing effects of UWOC channels, namely absorption, scattering and turbulence. In order to mitigate turbulence-induced fading and therefore to improve the system performance we use spatial diversity, i.e., employment of multiple transmitting lasers and/or multiple receiving apertures. We assume symbol-by-symbol processing and equal gain combining (EGC) at the receiver. In addition to evaluating the exact BER, we also evaluate an upper bound on the system BER from the inter-symbol interference (ISI) viewpoint. Moreover, we use Gauss-Hermite quadrature formula and also approximate the sum of log-normal random variables with an equivalent random variable, to effectively compute the average BERs.
\section{Channel and System Model}
\subsection{Channel Description}
Propagation of light in underwater medium is under the influence of three impairing phenomena, namely absorption, scattering and turbulence. Absorption and scattering processes cause loss on the received optical signal. Also scattering of photons temporally spreads the received optical signals and therefore limits the data transmission rate through inducing ISI. In order to take into account absorption and scattering effects of the underwater channel, we simulate the channel impulse response by Monte Carlo simulation method \cite{tang2014impulse,gabriel2013monte,cox2012simulation}. This turbulence-free impulse response of the UWOC channel between any two nodes, $i$th and $j$th, is denoted by $h_{0,ij}(t)$.

On the other hand, turbulence effects of the channel can be characterized by a multiplicative fading coefficient, ${\tilde{h}}_{ij}$ \cite{navidpour2007ber,andrews2005laser,lee2004part}. For weak oceanic turbulence, the aforementioned fading coefficient can be modeled as a random variable with log-normal probability density function (PDF) \cite{gerccekciouglu2014bit,yi2015underwater} as;
\begin{equation} \label{eq1}
f_{{\tilde{h}}_{ij}}\!({\tilde{h}}_{ij} )=\frac{1}{2{{\tilde{h}}_{ij}}\sqrt{2\pi {\sigma }^2_{X_{ij}}}}{\rm exp}\!\left(\!\!-\frac{{\left({{\rm ln}  ({{\tilde{h}}_{ij}})\ }\!\!\!-\!2{\mu }_{X_{ij}}\right)}^2}{8{\sigma }^2_{X_{ij}}}\!\right)\!\!,
 \end{equation}
  where  ${\mu }_{X_{ij}}$ and ${\sigma }^{2}_{X_{ij}}$ are mean and variance of the Gaussian distributed log-amplitude factor $X_{ij}=\frac{1}{2}{\rm ln}({{\tilde{h}}_{ij}})$. Therefore, the aggregated impulse response of the channel between any $i$th and $j$th nodes can be summarized as $h_{i,j}(t)={{\tilde{h}}_{ij}}h_{0,ij}(t)$. To insure that fading only makes fluctuations on the received optical signal, we should normalize fading coefficients as $\E[{\tilde{h}}_{ij}]=1$, which implies that ${\mu }_{X_{ij}}=-{\sigma }^{2}_{X_{ij}}$. 
  \subsection{System Model}
  Consider an UWOC system with $M$ transmitting lasers and $N$ receiving apertures. We assume on-off keying (OOK) modulation, i.e., the transmitter transmits each bit ``$1$" with pulse shape $P(t)$ and is off during transmission of data bit ``$0$". Hence, the total transmitted signal can be defined as $S(t)=\sum_{k=-\infty}^{\infty}b_kP(t-kT_b)$, where $b_k\in \left\{0,1\right\}$ is the $k$th time slot transmitted data bit and $T_b$ is the bit duration time. In the case of transmitter diversity, all the transmitters transmit the same data bit $b_k$ on their $k$th time slot. Therefore, the transmitted signal of the $i$th transmitter can be described as $S_i(t)=\sum_{k=-\infty}^{\infty}b_kP_i(t-kT_b)$, where $\sum_{i=1}^{M}P_i(t)=P(t)$, for the sake of fairness.
  
  Each $i$th transmitter, $TX_i$ is pointed to one of the receivers. The other receivers also capture the transmitted signal of $TX_i$ due to multiple scattering of photons under water. In other words, the transmitted signal of $TX_i$, $S_i(t)$ passes through channel with impulse response $h_{i,j}(t)$ to reach the $j$th receiver, $RX_j$. Therefore, the received optical signal from $TX_i$ to the $j$th receiver can be determined as;
  \begin{align} \label{eq2}
y_{i,j}(t)=S_i(t)*h_{i,j}(t)=\sum_{k=-\infty}^{\infty}b_k{\tilde{h}}_{ij}\Gamma_{i,j}(t-kT_b),
  \end{align}
in which $\Gamma_{i,j}(t)=P_i(t)*h_{0,ij}(t)$ and $*$ denotes convolution operation. Furthermore, $RX_j$ receives the transmitted signal of all the transmitters. Hence, we can express the total received optical signal of $RX_j$ as;
\begin{align} \label{eq3}
y_{j}(t)=\sum_{i=1}^{M}y_{i,j}(t)=\sum_{i=1}^{M}\sum_{k=-\infty}^{\infty}b_k{\tilde{h}}_{ij}\Gamma_{i,j}(t-kT_b).
  \end{align}

At the receiver side various noise components, i.e., background light, dark current, thermal noise and signal-dependent shot noise all affect the system operation. Since these components are additive and independent of each other, we model them as an equivalent noise component with Gaussian distribution \cite{lee2004part}. We also assume that the signal-dependent shot noise is negligible and hence the noise variance is independent of the received optical signal (see Appendix A).
\section{BER Analysis}
In this section we calculate the BER of UWOC system for both single-input single-output (SISO) and MIMO configurations. We assume symbol-by-symbol processing at the receiver side, which is suboptimal in the presence of ISI \cite{einarsson2008principles}. In other words, the receiver integrates its output current over each $T_b$ seconds and then compares the result with an appropriate threshold to detect the received data bit. In this detection process, the availability of channel state information (CSI) is assumed for threshold calculation \cite{navidpour2007ber}.
\subsection{SISO UWOC Link}
In SISO scheme, the $0$th time slot integrated current of the receiver output can be expressed as\footnote{The channel correlation time is on the order of $10^{-5}$ to $10^{-2}$ seconds \cite{tang2013temporal}. Therefore, the same fading coefficient is considered for all the consecutive bits in Eq. \eqref{eq4}};
\begin{align} \label{eq4}
r^{(b_0)}_{\rm SISO}=b_0\tilde{h}\gamma^{(s)}+\tilde{h}\sum_{k=-L}^{-1}b_k\gamma^{(k)}+v_{T_b},
\end{align}
where $\tilde{h}$ is the channel fading coefficient, $\gamma^{(s)}=\boldmath{R}\int_{0}^{T_b}\Gamma(t)dt$, $\boldmath{R}=\frac{\eta q}{hf}$ is the photodetector's responsivity, $\eta$ is the photodetector's quantum efficiency, $q=1.602\times10^{-19}$ C is electron charge, $h=6.626\times10^{-34}~{\rm J/s}$ is Planck's constant, $f$ is the optical source frequency and $L$ is the channel memory. Furthermore, $\gamma^{(k)}=\boldmath{R}\int_{0}^{T_b}\Gamma(t-kT_b)dt=\boldmath{R}\int_{-kT_b}^{-(k-1)T_b}\Gamma(t)dt$ interprets the ISI effect and $v_{T_b}$ is the receiver integrated noise component, which has a Gaussian distribution with mean zero and variance $\sigma^2_{T_b}$ \cite{lee2004part}. 

Assuming the availability of CSI, the receiver compares its integrated current over each $T_b$ seconds with an appropriate threshold, i.e., with $Th=\tilde{h}\gamma^{(s)}/2$. Therefore, the conditional probability of errors when bits ``$0$" and ``$1$" are transmitted, can be obtained respectively as;
\begin{align} \label{eq5}
P^{(\rm SISO)}_{be|0,\tilde{h},b_k}&=\Pr(r^{(b_0)}_{\rm SISO}\geq Th|b_0=0)\nonumber\\&=Q\left(\frac{\tilde{h}\left[\gamma^{(s)}/2-\sum_{k=-L}^{-1}b_k\gamma^{(k)}\right]}{\sigma_{T_b}}\right),
\end{align}
\begin{align} \label{eq6}
P^{(\rm SISO)}_{be|1,\tilde{h},b_k}&=\Pr(r^{(b_0)}_{\rm SISO}\leq Th|b_0=1)\nonumber\\&=Q\left(\frac{\tilde{h}\left[\gamma^{(s)}/2+\sum_{k=-L}^{-1}b_k\gamma^{(k)}\right]}{\sigma_{T_b}}\right),
\end{align}
where $Q\left(x\right)=({1}/{\sqrt{2\pi }})\int^{\infty }_x{{\rm exp}({-{y^2}/{2}})}dy$ is the Gaussian-Q function. The final BER can be obtained by averaging the conditional BER $P^{(\rm SISO)}_{be|\tilde{h},b_k}=\frac{1}{2}P^{(\rm SISO)}_{be|0,\tilde{h},b_k}+\frac{1}{2}P^{(\rm SISO)}_{be|1,\tilde{h},b_k}$, over fading coefficient $\tilde{h}$ and all $2^L$ possible data sequences for $b_k$s, as follows;
\begin{align}\label{eq7}
P^{(\rm SISO)}_{be}=\frac{1}{2^L}\sum_{b_k}\int_{0}^{\infty}P^{(\rm SISO)}_{be|\tilde{h},b_k}f_{\tilde{h}}(\tilde{h})d\tilde{h}.
\end{align}

The form of Eqs. \eqref{eq5} and \eqref{eq6} suggests an upper bound on the system BER, from the ISI point of view. In other words, $b_{k\neq0}=1$ maximizes Eq. \eqref{eq5}, while Eq. \eqref{eq6} has its maximum value for $b_{k\neq0}=0$. Indeed, when data bit ``$0$" is sent the worst effect of ISI occurs when all the surrounding bits are ``$1$" (i.e., when $b_{k\neq0}=1$), and vice versa \cite{einarsson2008principles}. Regarding to these special sequences, the upper bound on the BER of SISO-UWOC system can be evaluated as;
\begin{align}\label{eq8}
P^{(\rm SISO)}_{be,upper}=\frac{1}{2}\int_{0}^{\infty}\Bigg[& Q\left(\frac{\tilde{h}\left[\gamma^{(s)}/2-\sum_{k=-L}^{-1}\gamma^{(k)}\right]}{\sigma_{T_b}}\right)\nonumber\\&+Q\left(\frac{\tilde{h}\gamma^{(s)}}{2\sigma_{T_b}}\right)\Bigg]f_{\tilde{h}}(\tilde{h})d\tilde{h}.
\end{align}

The averaging in Eqs. \eqref{eq7} and \eqref{eq8} over fading coefficient, involves integrals of the form $\int_{0}^{\infty}Q(C\tilde{h})f_{\tilde{h}}(\tilde{h})d\tilde{h}$, where $C$ is a constant, e.g., $C=\gamma^{(s)}/2\sigma_{T_b}$ in second integral of Eq. \eqref{eq8}. Such integrals can be calculated by Gauss-Hermite quadrature formula [15, Eq. (25.4.46)] as follows;
\begin{align} \label{eq9}
&\int_{0}^{\infty}Q(C\tilde{h})f_{\tilde{h}}(\tilde{h})d\tilde{h}\nonumber\\
&=\int_{-\infty}^{\infty}Q(Ce^{2x})\frac{1}{\sqrt{2\pi\sigma^2_X}}\exp\left(-\frac{(x-\mu_X)^2}{2\sigma^2_X}\right)dx\nonumber\\
&\approx \frac{1}{\sqrt{\pi}}\sum_{q=1}^{U}w_qQ\left(C \exp\left(2x_q\sqrt{2\sigma^2_X}+2\mu_X\right)\right),
\end{align}
in which $U$ is the order of approximation, $w_q,~q=1,2,...,U$, are weights of $U$th order approximation and $x_q$ is the $q$th zero of the $U$th-order Hermite polynomial, $H_U(x)$ \cite{navidpour2007ber,abramowitz1970handbook}.
\subsection{MIMO UWOC Link}
Assume a multiple-input multiple-output UWOC system with equal gain combiner (EGC). The integrated current of the receiver output can be expressed as;
\begin{align} \label{eq10}
r^{(b_0)}_{\rm MIMO}=b_0\sum_{j=1}^{N}\sum_{i=1}^{M}{\tilde{h}}_{ij}\gamma^{(s)}_{i,j}+\!\sum_{j=1}^{N}\sum_{i=1}^{M}{\tilde{h}}_{ij}\!\!\!\!\sum_{k=-L_{ij}}^{-1}\!\!\!b_k\gamma^{(k)}_{i,j}+v^{(N)}_{T_b},
\end{align}
where $\gamma^{(s)}_{i,j}=\boldmath{R}\int_{0}^{T_b}\Gamma_{i,j}(t)dt$, $\gamma^{(k)}_{i,j}=\boldmath{R}\int_{0}^{T_b}\Gamma_{i,j}(t-kT_b)dt=\boldmath{R}\int_{-kT_b}^{-(k-1)T_b}\Gamma_{i,j}(t)dt$ and $v^{(N)}_{T_b}$ is the integrated combined noise component, which has a Gaussian distribution with mean zero and variance $N\sigma^2_{T_b}$.\footnote{Note that the received background power is proportional to the receiver aperture area. However based on Appendix A, the background noise has negligible contribution on the total noise of the receiver. Moreover, each receiver has its distinct dark current and thermal noise. Hence, the noise variance in MIMO scheme is $N$ times of that in SISO case.}

Based on Eq. \eqref{eq10} and availability of CSI, in MIMO scheme the receiver selects the threshold value as $Th_{\rm MIMO}=\sum_{j=1}^{N}\sum_{i=1}^{M}{\tilde{h}}_{ij}\gamma^{(s)}_{i,j}/2$. Therefore, pursuing similar procedures as Section III-A results into the following equation for conditional BER.
\begin{align} \label{eq11}
& P^{(\rm MIMO)}_{be|b_0,{\boldsymbol{\bar{H}}},b_k}=\nonumber\\ & Q\!\!\left(\!\frac{\!\sum_{j=\!1}^{N}\!\sum_{i=\!1}^{M}\!{\tilde{h}}_{ij}\gamma^{(s)}_{i,j}\!-\!(\!-\!1)^{b_0}\!\!\sum_{j=\!1}^{N}\!\sum_{i=\!1}^{M}\!{\tilde{h}}_{ij}\!\sum_{k=-L_{ij}}^{-1}\!\!\!2b_k\!\gamma^{(k)}_{i,j}}{2\sqrt{N}\sigma_{T_b}}\!\!\right)\!\!,
\end{align}
in which ${\boldsymbol{\bar{H}}}=\{{\tilde{h}}_{11},{\tilde{h}}_{12},...,{\tilde{h}}_{MN}\}$ is the fading coefficients' vector. Assume the maximum channel memory to be $L_{\rm max}=\max\{L_{11},L_{12},...,L_{MN}\}$, then the average BER of MIMO-UWOC system can be obtained by averaging over ${\boldsymbol{\bar{H}}}$ (through $M\times N$-dimensional integral) as well as averaging over all $2^{L_{\rm max}}$ sequences for $b_k$s;
\begin{align}\label{eq13}
& P^{(\rm MIMO)}_{be}=\nonumber\\
&\frac{1}{2^{L_{\rm max}}}\sum_{b_k}\int_{{\boldsymbol{\bar{H}}}}\frac{1}{2}\left[P^{(\rm MIMO)}_{be|1,{\boldsymbol{\bar{H}}},b_k}+P^{(\rm MIMO)}_{be|0,{\boldsymbol{\bar{H}}},b_k}\right]f_{{\boldsymbol{\bar{H}}}}({\boldsymbol{\bar{H}}})d{\boldsymbol{\bar{H}}},
\end{align}
where $f_{{\boldsymbol{\bar{H}}}}({\boldsymbol{\bar{H}}})$ is the joint PDF of fading coefficients in ${\boldsymbol{\bar{H}}}$.

Similar to Section III-A, the upper bound on the BER of MIMO-UWOC system can be evaluated by considering the transmitted data sequences as $b_{k\neq0}=1$ for $b_0=0$ and $b_{k\neq1}=0$ for $b_0=1$. Moreover, similar to Eq. \eqref{eq9} the $M\times N$-dimensional integral in Eq. \eqref{eq13} can be approximated by $M\times N$-dimensional series, using Gauss-Hermite quadrature formula.

It's worth noting that the sum of random variables in Eq. \eqref{eq11} can be effectively approximated by an equivalent random variable, using moment matching method \cite{fenton1960sum}. In other words, we can reformulate the numerator of Eq. \eqref{eq11} as $\beta^{(b_0)}=\sum_{j=1}^{N}\sum_{i=1}^{M}{\boldmath{G}}^{(b_0)}_{i,j}{\tilde{h}}_{ij}$, i.e., the weighted sum of $M\times N$ random variables. The weight coefficients are defined as ${\boldmath{G}}^{(b_0)}_{i,j}=\gamma^{(s)}_{i,j}+(-1)^{b_0+1}\sum_{k=-L_{ij}}^{-1}2b_k\gamma^{(k)}_{i,j}$. In the special case of log-normal distribution for fading coefficients, $\beta^{(b_0)}$ can be approximated with an equivalent log-normal random variable as $\beta^{(b_0)}\approx\alpha^{(b_0)}=\exp(2z^{(b_0)})$, with log-amplitude mean $\mu_{z^{(b_0)}}$ and variance $\sigma^2_{z^{(b_0)}}$ of \cite{safari2008relay};
\begin{align} \label{eq14}
\mu_{z^{(b_0)}}=\frac{1}{2}{\rm ln}\bigg(\sum_{j=1}^N\sum_{i=1}^M{\boldmath{G}}^{(b_0)}_{i,j}\bigg)-\sigma^2_{z^{(b_0)}},
\end{align}
\begin{align} \label{eq15}
\sigma^2_{z^{(b_0)}}\!\!=\!\frac{1}{4}{\rm ln}\!\left(\!1\!+\!\frac{\sum_{j=1}^N\sum_{i=1}^M\!\!\left({\boldmath{G}}^{(b_0)}_{i,j}\right)^2\!\!\left(e^{4\sigma^2_{X_{ij}}}-1\right)}{\left(\sum_{j=1}^N\sum_{i=1}^M{\boldmath{G}}^{(b_0)}_{i,j}\right)^2}\right).
\end{align}
Then averaging over fading coefficients reduces to one-dimensional integral of
\begin{align} \label{eq16}
P^{(\rm MIMO)}_{be|b_0,b_k}\approx\int_{0}^{\infty}Q\left(\frac{\alpha^{(b_0)}}{2\sqrt{N}\sigma_{T_b}}\right) f_{\alpha^{(b_0)}}(\alpha^{(b_0)})d\alpha^{(b_0)},
\end{align} 
which can be effectively calculated using Eq. \eqref{eq9}.
\section{Numerical Results}
In this section we present the numerical results for BER of MIMO-UWOC systems. The system is assumed to be established in coastal water which has attenuation, absorption and extinction coefficients of $a={0.179}$ \si{m^{-1}}, $b={0.219}$ \si{m^{-1}} and $c={0.398}$ \si{m^{-1}}, respectively \cite{mobley1994light}. Further, we assume rectangular pulse shape for transmitted data bits, i.e., $P(t)=P\Pi \left(\frac{t-{T_b}/{2}}{T_b}\right)$, where $P$ is the total transmitted power per bit ``$1$" and $\Pi(t)$ is a rectangular pulse with unit amplitude in the interval $[-1/2,1/2]$. Moreover, we assume the same transmitted power of $P/M$ for all the transmitters and the same receiving aperture area of $A_r/N$ for all the receivers, where $A_r$ is the total aperture area of the receiver.
Table I shows some of the parameters which we have assumed for the channel fading-free impulse response simulation and the noise characterization (refer to Appendix A for further descriptions on the noise components characterization).

\begin{table}
  \centering
\caption{Some of the important parameters for channel simulation and noise characterization.}
\begin{tabular}{||p{1.6in}|>{\centering\arraybackslash}p{0.3in}|>{\centering\arraybackslash}p{0.8in}||} \hline
   Coefficient & Symbol & Value\\ [0.2ex] 
   \hline \hline
  Half angle field of view & FOV &  ${40}^0$ \\ \hline 
     Receiver aperture diameter & $D_0$ & $20$ cm \\ \hline 
     Source wavelength & $\lambda $ & $532$ nm \\ \hline 
     Water refractive index & $n$ & $1.331$ \\ \hline 
     Source full beam divergence & $\theta_{div}$ & $0.02^0$ \\ \hline 
     Photon weight threshold at the receiver & $w_{th}$ & ${10}^{-6}$ \\ \hline 
     Quantum efficiency & $\eta $ & $0.8$ \\ \hline 
         Electronic bandwidth & $B$ & $10$ GHz \\ \hline
Optical filter bandwidth & $\Delta\lambda$ & $10$~nm \\ \hline
         Equivalent temperature & $T_e$ & $290$ K \\ \hline 
         Load resistance & $R_L$ & $100$ $\Omega$ \\ \hline 
         Dark current & $I_{dc}$ & $1.226\times {10}^{-9}$ A \\ \hline
      Received background power &$P_{BG}$& $6.34\times10^{-11}$ W \\ \hline
  \end{tabular}
  \end{table}  

\begin{figure}[t]
\centering
\includegraphics[width=3.4in,height=2.2in]{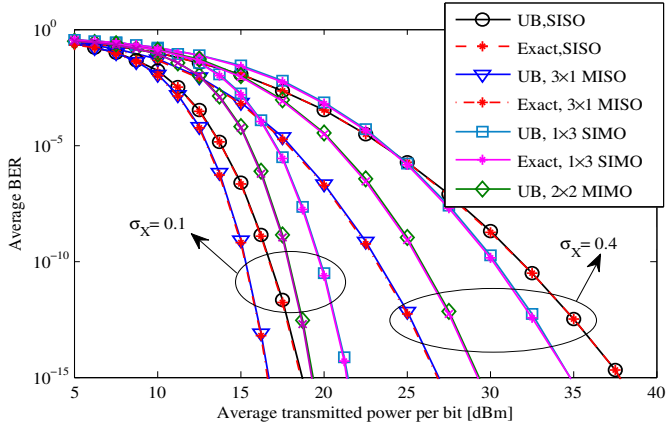}
\caption{Upper bound (UB) and exact BER of a $25$ m coastal water link with different configurations. $R_b=1$ Gbps, ${\sigma }_X=0.1$ and $0.4$.}
\label{Fig1}
       \end{figure}

Fig. \ref{Fig1} depicts the upper bound and exact BER of a $25$ m coastal water link with different configurations and data rate of $R_b=1$ Gbps. As it can be seen, in all the configurations upper bound curves have excellent matches with the exact BER curves. Moreover, in a relatively strong turbulent channel, e.g., ${\sigma }_X=0.4$, spatial diversity (especially at the transmitter side) can introduce a noticeable performance improvement, e.g., $6$ dB and $9$ dB at the BER of $10^{-12}$, using two and three transmitters, respectively. This achievement relatively vanishes in very weak turbulence regimes, e.g., ${\sigma }_X=0.1$, where fading has a negligible effect on the system performance. Furthermore, this figure shows that the transmitter diversity performs better than the receiver diversity, due to less noise power and larger aperture area.
  
       \begin{figure}[t]
 \centering
\includegraphics[width=3.4in,height=2.2in]{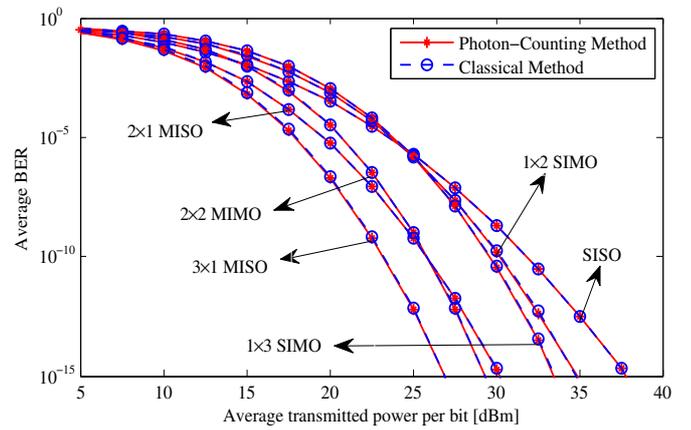}
\caption{Accuracy of the obtained expressions; upper bound BER of a $25$ m coastal water link with different configurations, $R_b=1$ Gbps and ${\sigma }_X=0.4$, obtained using photon-counting method and classical approach of this paper.}
\label{Fig2}
  \end{figure}
\begin{figure}[t]
 \centering
\includegraphics[width=3.4in,height=2.2in]{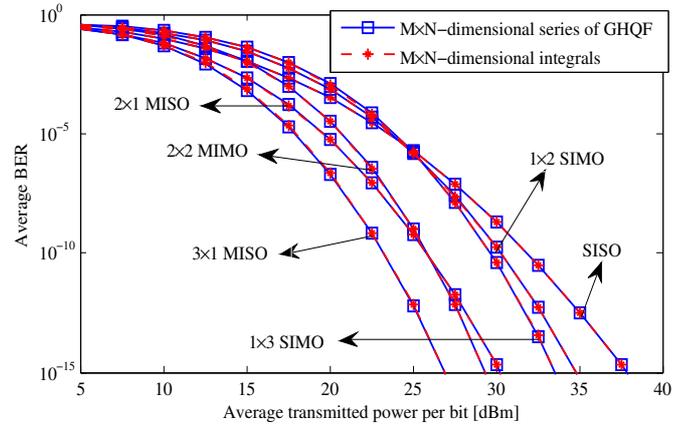}
\caption{Comparing the results of $M\times N$-dimensional series of Gauss-Hermite quadrature formula (GHQF) and $M\times N$-dimensional integrals of Eqs. \eqref{eq11}, \eqref{eq13} in evaluating the upper bound BER of a $25$ m coastal water link with different configurations, $R_b=1$ Gbps and ${\sigma }_X=0.4$.}
\label{Fig3}
 \end{figure}
 \begin{figure}[t]
\centering
\includegraphics[width=3.4in,height=2.2in]{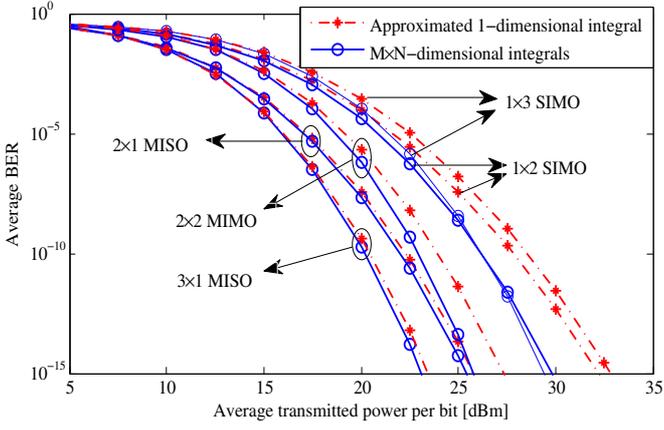}
\caption{Comparing the results of one-dimensional integral of Eq. \eqref{eq16} and $M\times N$-dimensional integrals of Eqs. \eqref{eq11}, \eqref{eq13} in evaluating the upper bound BER of a $25$ m coastal water link with different configurations, $R_b=1$ Gbps and ${\sigma }_X=0.4$.}
\label{Fig4}
\end{figure}
 
  In Figs. \ref{Fig2}, \ref{Fig3} and \ref{Fig4} the upper bound BER of a $25$ m coastal water link with ${\sigma }_X=0.4$ and $R_b=1$ Gbps is illustrated for different configurations. In particular, Fig. \ref{Fig2} compares the results of $M\times N$-dimensional integrals of Eqs. \eqref{eq11}, \eqref{eq13} with the photon-counting method results \cite{jamali2015performance-mimo}. As it is obvious, excellent matches between these curves confirm the accuracy of the derived expressions in this paper.
  It is worth noting that SIMO schemes, which compensate for fading impairments (at high SNR regimes) can not outperform SISO performance, except at low BERs. This is mainly due to that each receiver in SIMO scheme has $N$ times less aperture area than SISO receiver and also SIMO system has $N$ times larger noise contribution than SISO scheme.
  
   In Fig. \ref{Fig3} we applied Gauss-Hermite quadrature formula (GHQF) to approximate the $M\times N$-dimensional integrals of Eqs. \eqref{eq11}, \eqref{eq13} with $M\times N$-dimensional series, using Eq. \eqref{eq9}. The order of approximation is assumed to be $U=30$. Obviously, GHQF can effectively compute the $M\times N$-dimensional integrals (even with less than $U=30$ points for each integral). In Fig. \ref{Fig4} we used Eqs. \eqref{eq14}-\eqref{eq16} to approximate the weighted sum of log-normal random variables in Eq. \eqref{eq11} with an equivalent log-normal random variable. As it can bee seen, this approximation provides an excellent estimate of the BER of UWOC system with transmitter diversity. However, the discrepancy increases for the case of receiver diversity. In other words, in the case of transmitter diversity all the transmitters are pointed to a single receiver and therefore, all the links have the same weight coefficient of $G^{(b_0)}_{i,1}=G^{(b_0)}_{\rm MISO}$. Hence, BER expression of MISO-UWOC can be estimated using approximation of unweighted sum of log-normal random variables \cite{navidpour2007ber}.

\section{Conclusion}
In this paper we analytically calculated the BER of a MIMO-UWOC system with equal gain combining and symbol-by-symbol processing. Our analytical treatment included all the disturbing effects of the UWOC channels, i.e., absorption, scattering and fading. we obtained both the exact and upper bound BER expressions. Also we used Gauss-Hermite quadrature formula to more effectively calculate the averaging integrals with finite series. Moreover, we approximated the weighted sum of log-normal random variables with an equivalent log-normal random variable to reduce $M\times N$-dimensional integrals of averaging (over fading coefficients) to one-dimensional integrals. Our analytical results showed well match between the exact and upper bound BERs and also the results of Gauss-Hermite quadrature formula. Furthermore, we observed that MIMO transmission can introduce a noteworthy performance improvement in relatively high turbulent UWOC channels. However, we assumed log-normal distribution for fading statistics, we should emphasize that most of our derivations are applicable for any fading distribution.
\appendices
\section{Negligibility of Signal-Dependent Shot Noise}
In this appendix we verify the validity of assumption that ``the signal-dependent shot noise is negligible with respect to the other noise components". To do that, we should satisfy the inequality $\sigma^2_{ss}\ll \sigma^2_{BG}+\sigma^2_{DC}+\sigma^2_{TH}$, where $\sigma^2_{ss}$, $\sigma^2_{BG}$, $\sigma^2_{DC}$ and $\sigma^2_{TH}$ are respectively the current variance of the Gaussian distributed signal-dependent shot noise, background light, dark current and thermal noise \cite{jaruwatanadilok2008underwater}. In order to verify the validity of the above mentioned assumption we should satisfy the following inequality;
\begin{align}
2\frac{\eta P^{(rec)}_s}{hf}q^2B &\ll 2\frac{\eta P_{BG}}{hf}q^2B +2qI_{dc}B+\frac{4KT_eB}{R_L} \label{eq_ap1}\\
\Rightarrow P^{(rec)}_s &\ll P_{BG}+\frac{I_{dc}}{\eta q}hf+\frac{2KT_e}{\eta q^2R_L}hf. \label{eq_ap2}
\end{align}
With respect to the parameters in Table I, Eq. \eqref{eq_ap2} simplifies to $P^{(rec)}_s \ll 6.34\times10^{-11}+2.688\times10^{-9}+1.097\times10^{-3}~{\rm W}\approx 1~{\rm mW}$. Here, the background noise power is calculated in a similar procedure to \cite{jaruwatanadilok2008underwater}. Some of the assumed parameters are shown in Table I and the other parameters are exactly the same as those are in \cite{jaruwatanadilok2008underwater}.

To gain more insight on validity of the aforementioned assumption let's to evaluate the BER of an UWOC system with $P^{(rec)}_s \ll 1~{\rm mW}$ and high ISI, using Gaussian approximation \cite{einarsson2008principles}. Assume an UWOC system with $P^{(rec)}_s=10^{-4}~{\rm mW}$ received power for transmitted bit ``$1$", which implies to the mean photoelectron counts of $m_1=\frac{\eta P^{(rec)}_s }{hf}T_b\approx 2.85\times 10^{5}$, for $T_b=1~{\rm ns}$. Also assume the photoelectrons count to be $m_0=m_1/2=1.425\times 10^{5}$, conditioned on transmission of bit ``$0$" ($m_0=m_1/2$ relates a channel with high ISI). With respect to the parameters of Table I, (count) noise variance can be obtained as $\sigma^2_{m}\approx \sigma^2_{m,TH}=\frac{2KT_eT_b}{R_Lq^2}\approx 3.12\times10^6$ \cite{jazayerifar2006atmospheric}. Using Gaussian approximation, the BER of an UWOC system with the above parameters for $m_0$, $m_1$ and $\sigma^2_m$ can be obtained as;
\begin{align} \label{eq_ap3}
P_{be} \approx Q\left(\frac{m_1-m_0}{\sqrt{m_1+\sigma^2_m}+\sqrt{m_0+\sigma^2_m}}\right)\approx Q(40)\approx 0.
\end{align} 
Therefore, the aforementioned assumption is often valid for a wide range of BERs. Note that for typical values of BER, $P_s^{(rec)}$ has smaller  values than $10^{-4}$ W, and hence the above assumption is more valid.




\bibliographystyle{IEEEtran}
\bibliography{IEEEabrv}
\end{document}